# Understanding Code Patterns – Analysis, Interpretation & Measurement

Jitesh Dundas, *Member, IACSIT*

*Abstract*—This research paper aims to find, analyze and understand code patterns in any software system and measure its quality by defining standards and proposing a formula for the same. Every code that is written can be divided into different code segments, each having its own impact on the overall system. We can analyze these code segments to get the code quality. The measures used in this paper include Lines of Code, Number of calls made by a module, Execution time, the system knowledge of user and developers, the use of generalization, inheritance, reusability and other object-oriented concepts. The entire software code is divided into code snippets, based on the logic that they implement. Each of these code snippets has an impact. This measure is called Impact Factor and is valued by the software developer and/or other system stakeholders.

Efficiency = (Code Area / Execution Time) * Qr

*Index Terms*— code area, code patterns, efficiency, lines of code, pattern recognition, software quality

## I. INTRODUCTION

The aim of this paper is to find, analyze and understand code patterns in any system and to measure system and developer's behavior and quality by defining standards and proposing a formula for the same.

Every code that is written can be divided into different patterns that reflect the pros and cons of the same. We can also do the analysis of the code and understand the quality[1] of code and the expected behavior.

The measures used for this paper are:
1) Lines Of Code-Length of code shows the quality of code.
   A good programmer will code faster, better and more accurate than those with lower levels of coding skills.
2) Number of calls made by a module.
3) Execution time.
4) System knowledge of user and developers.
5) Use of generalization, inheritance, reusability and other object-oriented concepts.

For e.g.) this code snippet mentioned below:-
a=0
If (a == 3)
{
   a = a + 1

Else
{
   a = a - 1
}

Each of these lines has an impact due to the logical importance in the software i.e. the Else cannot be as important as the line a = a - 1 or a = a + 1. We actually dilute the impact and accuracy of the actual quality measurement. Each of the code snippets carry an impact, all of which accumulate to give the actual code quality. In the diagram shown in the last page, the loops or recursive calls have not been shown on the lines at any point in the program flow. Thus, they have been considered as a single point. It can be shown as a circle sitting on the lines of the program flow. They also have flow moving in the forward direction. In cases where the flow is moving backwards or in an asynchronous or structured manner, the software has been observed to be of low quality.

As in Fig-2), The Code Area will consider the number of lines covered within each loop. Two things are significant here:-
1) The end point or maximum value of the counter till which the loop will execute.
2) The number of lines executed within the condition.

For e.g.) if the code in (C++ or Lotus script languages), the loop will be:-

For (i=0; i < 100; i++)
{
   Printf ("Hello Everyone!");
   Printf ("I am XYZ!");
}

The total code area here will be (100 * 2 = 200). Thus, the same applies for nested loops i.e. loop within loops.

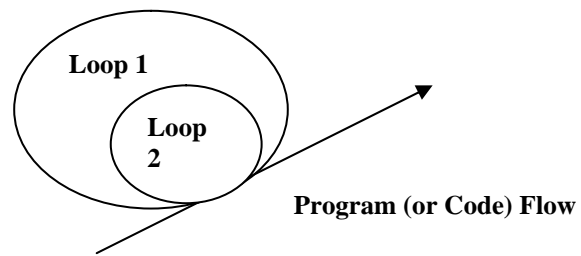

Fig. 1 Nested Loop representation

Jitesh Dundas is working as a research associate with Edencore Technologies(http://www.myspace.com/jordan_digital). He is also a Web Developer with JR Technologies, Mumbai-400076, India. He can be reached at (+91-9860925706; e-mail: jbdundas@gmail.com). This work was not supported by any institution financially.

**Also the line 'k' can be denoted by the equation:-**

$$k = \sum_{l=1}^{n} [\ldots ((1 * m-1) * m-1) \ldots m-1)]_{l=1}^{n} \quad (1)$$

Code Segment Area is the term coined in this paper to quantify the amount of space that is consumed or occupied by the system. It is expressed in terms of Lines of Code (LOC).

Code Segment Execution Time is defined as the time taken by the system function to execute its code once the call is made to it. It is expressed in seconds.

## II. FORMULAE TO ESTIMATE SOFTWARE QUALITY USING LOC

Quality is a major factor in object-oriented software development [5]. Thus, measuring quality of logic implementation is very important. The entire software code is divided into code snippets, based on the logic that they implement. Each of these code snippets has an impact. This measure is called Impact Factor and is valued by the software developer and/or other system stakeholders.

1) Simple Lines

These lines include simple declarations, assignment statements, and initiation and termination statements.

a) If for declaration statements = 1
b) If for initiation and termination statements = 2
c) Other assignment statements =
   simple = 3-4
   complex = 5-7
   expressions (very complex) = 8-10

These measures can be changed by the developers as their estimates.

Lines Of Code (LOC) = SL + CL + LL + EL where,   (2)

SL = Simple Lines
CL = Condition-Based Lines
LL = Looped-Based Lines
EL = Exception-Handling Lines

Here,

SL = number of simple lines. It can be manually counted.
EL = Number of lines written to handle exception.

If function is called, then multiply as in section-A. Finally multiply the value by the number of exception blocks present in the code segment, say n. We may need to find the impact of the exception segments by adding the impacts of the condition, loop and simple line segments.

$$LL = (n * m \ldots (t * l)) * I_{fi} \quad (3)$$

where n and m are the number of loops and l is the number of lines to be counted. Finally multiply the value by the number of exception blocks present in the code segment, say n. Please note that for any nested conditions, we can replace simple line variable with the nested condition's expression (i.e. $Cl_{fi}$ variable for condition-based lines' impact).

CL = For 'h' number of conditions present in a condition block,
 Avg. number of lines of code = $(1 / h) * l * m$

$$CL_{fi} = \sum_{I=1}^{n} CL_i * I_{fi} \quad (4)$$

Where m = Number of lines executed in that condition.
h = Average Success Ratio of conditions being executed
l = Number of conditions present in the condition block.

Finally multiply the value by the number of exception blocks present in the code segment, say n.

Please note that for any nested loops within the condition, we can replace simple line variable with the nested loop's expression (i.e. LL variable for loop-based lines' impact). LOC here is the area of that segment only. Many such segments (Total N code segments make up the software) are present in the software.

Thus, we can imply that:
    Avg. Code Segment Size = LOC   (5)

    $L_c = n_1*SL + n_2*CL + n_3*LL + n_4*EL$

Where $n_1, n_2, n_3, n_4$ are the number of code segments for Simple Lines, Condition-Based Lines, Looped based lines and exception-handling lines.

Code Area = N * Avg. Code Segment Size (LOC)

Execution Time (in seconds) = N * Avg. Code Segment Execution Time.   (6)

$Q_r$ = Quality Quotient (on a scale of 1 to 10 given by user). This will include the following aspects:-
1) Security
2) Execution Time
3) User-Friendliness
4) Other software metrics for measuring software quality.
5) Optimum Programming Environment Selection

Other possible metrics quality evaluation criteria[6] include reusability, testability/maintainability and understandability.

Give a measure of 0, 1 or 2 based on the perception of the user of how much is that quality attribute present in the software code[2]. Thus, final answer would be out of 10. For e.g.) 1/10 till 10/10.

Efficiency = (Code Area / Execution Time) * $Q_r$   (7)

The higher the efficiency, the better the software quality is. Generally the code can be expected to have at least 75% expected efficiency to be considered of good quality. This percentage is calculated by dividing our obtained 'Efficiency' measure against the efficiency of 100, 000LOC at a quality rate of 7.5.

III. EXPLANATION

COCOMO[8] is one of the widely accepted models in software estimation. The Constructive Cost Model (COCOMO) is an algorithmic Software Cost Estimation Model for estimating effort, cost, and schedule for software projects.

Every code that is written can be divided into different patterns that reflect the pros and cons of the same. We can also do the analysis of the code and understand the quality of software[7].The quality of code also reflects the speed and intelligence of the developer (or even the attitude or aim of the team) The code written in short-term reflects that the team was interested in just finishing the applications or that they did not have enough time to do it (most common reason found till now).The coding skills of the developer were not up to the level of that system's development requirements.

Too much time taken in making routine simple code (applications reflect any or the following things) means that

The coding skills of developer are of low level or that too many problems faced while development e.g.) requirement change or technical issues.

The following are the measures used in the paper for measuring software quality:

1) Lines Of Code- Length of code shows the quality of code. A good programmer will code faster, better and more accurately than those with lower levels of coding skills.

2) No. Of calls made by a module.

3) Execution time.

4) System knowledge of user and developers.

5) Use of generalization, inheritance, reusability and other object-oriented concepts.

6) Security and

7) Compatibility between scripting languages and databases, connectivity issues, etc.

Generally there are scripting languages that are used for writing code. In such cases, there should be interoperability or common functions that allow understanding of their scripts between each other.

The code and design has to be structured and written so as to be in sync with the flow of information. It should allow the user to move smoothly to perform his work.

It must let him do 3 things:-

1) To let user move to next task as soon as current Task is done.
2) To let user get assistance or help as and when needed.
3) To exit or move back if and when needed.

There are also some behavioral aspects involved to this method. The type of code that has been written in the software can be classified into various levels. These levels have their own interpretations. They are-

*1) Level-1(Free-Flow)*

Here the code segments are continuous and have smooth execution of function calls. There are no uncontrolled loops with only 1(or maximum 2) termination choices to handle failures. Here, every feature or code syntax is used at the right place at the right time. Entire code is divided in segments, executed in order in the form of functions or script libraries. There is high use of object-oriented concepts in the code.

Each code segment is properly commented in easily-understandable language. They are also error-free, logically and syntactically. The user interfaces are easy to use, well-connected by links, user-friendly and giving the exact reflection of the organization or purpose which it represents. There is proper help and guidance available with contact information. The policies and beliefs of the organization are taken into consideration and shown favorably in the user-interfaces. The interfaces will also show what the organization believes in and does not. The effort made by the user to interact with the system is minimum and the system does everything for the user. The system takes care of everything. The security of the system is very high, ranging from the field-level restrictions to the network-level restrictions with user-based customization. The user-interfaces will be intelligent with the logic to decide the system's behavior based upon the user that accesses the system. This includes managing user's behavioral knowledge like preferences, frequently used links, likes and dislikes and expectations. The intelligent and self-learning system will behave and react as per the type of user interacting with the system. This could also include promoting brands, advertisements and special offers. Such systems have a range of 8.5 – 10 on a scale of 1-10. This category has systems that are closest to being perfect.

*2) Level-2*

Here the code segments are continuous and have smooth execution of function calls. There are uncontrolled loops with 2-3 termination choices to handle failures. Here, every feature or code syntax is used properly, but not always at the right place at the right time. Entire code is divided in segments, but not guaranteed to execute in order (using functions or script libraries). There is average use of object-oriented concepts in the code with duplication and redundancy being present at minor levels. Each code segment is commented in not- so-easy and little complicated language. They are also error-prone at minor levels, logically and syntactically. The user interfaces are difficult to use, well-connected by links, not so user-friendly and giving the basic reflection of the organization or purpose which it represents. There is no proper help and average guidance available with contact information. The policies and beliefs of the organization are taken into consideration partially and shown in the user-interfaces without any favorable impacts. The interfaces will also show what the organization believes in and does not but without catching the interests of the user. The effort made by the user to interact with the system is high and the system does everything for the user based on this effort. The system takes partial care of user's requirements. The security of the system is high, ranging from the field-level restrictions to the network-level restrictions with user-based customization. However, security will not be free from malfunctions and thefts. The user-interfaces will be not being intelligent without the logic to decide the system's behavior based upon the user that accesses the system. This excludes managing user's behavioral knowledge like preferences, frequently used links, likes and dislikes and expectations. This descent and WYSIWYG system will behave and react as per the input of user interacting with the system. This excludes also promoting brands, advertisements and special offers. Only the information will be displayed irrespective of user-preferences. Such systems have a range of 6.5-8 on a scale of 1-10. This category has systems that are descent, providing information with average intelligence and basic results.

*3) Level-3*

Here the code segments are irregular and have rough execution of function calls. There is presence of duplicate or redundant code at a lot of places in the system. There are uncontrolled loops with 3-5 termination choices to handle failures. Here, every feature or code syntax is used, but rarely at the right place at the right time. Entire code is divided in segments, executed in disorder with low use of functions and script libraries. There is low use of object-oriented concepts in the code. Each code segment is badly- commented in complicated or difficult language. They are also highly error-prone, logically and syntactically. The user interfaces are difficult to use, not so-well-connected by links or without links, not user-friendly and giving the no reflection of the organization or purpose which it supposedly represents. There is no proper help and low- guidance available with basic contact information. The policies and beliefs of the organization are not taken into consideration and not shown in the user-interfaces. The interfaces will not show what the organization believes in and does not. The effort made by the user to interact with the system is high and the system does little for the user. The system takes little care of things without input at every stage. Reliability of the system is less. The security of the system is low, ranging from the field-level

The user interfaces are difficult to use, poorly-connected by links, not user-friendly and giving the opposite or wrong reflection of the organization or purpose which it represents. There is poor or no help and guidance available with contact information. The policies and beliefs of the organization are not taken into consideration and shown unfavorably in the user-interfaces. The interfaces will also not show what the organization believes in and does not. The effort made by the user to interact with the system is maximum and the system does little for the user. The system needs input and maintenance at every level. The security of the system is average or low, ranging from the field-level restrictions to the network-level restrictions with user-based customization. The user-interfaces will be dumb without the logic to decide the system's behavior based upon the user that accesses the system. This excludes managing user's behavioral knowledge.

restrictions to the network-level restrictions with little user-based customization. The user-interfaces will be without the logic to decide the system's behavior based upon the user that accesses the system. This excludes managing user's behavioral knowledge like preferences, frequently used links, likes and dislikes and expectations. This average and simple system will behave and react as per the user input and error-frequency affecting the system. This also excludes promoting brands, advertisements and special offers. Only simple information like notices, static data can be displayed. Repairing the poorly developed system will involve patchwork that can be expected to be time-consuming and costly. Such systems have a range of 4.5-6 on a scale of 1-10. This category has systems that are average, unreliable and error-prone with high maintenance. It is always good to remove or repair such systems.

*4) Level-4*

Here the code segments are unreliable, highly error-prone and have rough or no execution of function calls. There are uncontrolled loops with more than 5 termination choices to handle failures. Here, every feature or code syntax is badly placed and written. Entire code is continuous without segments or divided illogically in segments, executed in disorder with minimum or without functions or script libraries. There is very less use of object-oriented concepts in the code. Each code segment is poorly commented in difficult and complex language. They are also error-prone, logically and syntactically.

I. EXAMPLE

**Here is a working example of the method described in the above pages.**

We take the entire code [9] shown in Fig-9) and then divide into logically separate code segments, with each code segment implementing a separate logic of the program.

1) Simple Lines

Please note here that there is only logic being implemented here. These are simple comment lines and thus they all can be clubbed into one code segment. Next, we multiple the simple code lines into with the Impact factor, which we could measure as 5/10 or 0.5 (since these are simple lines with no assignments or expression calculations) Thus, the result we get is code segment impact. Here, it is 20 * 0.5 = 10.

In the same way, there can be different simple lines, which alone or in concurrent groups of lines, are implementing logic of the program. They can be measured in the same way as done above. The number of lines and the impact factor could be different. In the end, we add the impact of all such code segments for simple lines and add them together. Please note that we add them together here as this code is made up of simple lines. In the next example shown in the figure-3) below, we again get some comments in the first few lines. We just get the Impact as done earlier. Next, we see that there are Header Files. These can be treated as simple lines with a larger impact as they are the calls to library functions. Thus, we get this impact here for the 3 lines (impact factor for each line = 0.7) as 2.1 (as shown in the diagram). Next, we see that

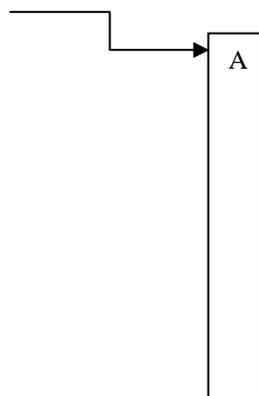

Fig. 2 Code Area Representation

there are functions defined in the program. We treat them as separate programs, and thus finding their respective impacts and then adding them to the impact of this program in the end. We can also do this step add the time of adding the impacts of all the programs in the software. Please see Figure-4) for further details. As these are all simple lines with high impact, the

Impact of each of them have been added together to give us the impact for all the lines here.

- Code Impact = 2.1 + 1 * 0.8 + 3 * 0.8 = 5.3

- Code Impact = 5.3

As shown in Figure-5), the If-Else condition here has only 2 cases: - if and else. Thus we get here
Condition-Based Lines (CL) Impact = No. Of Cases * Statements * Impact

- CL = ½ * X

- Where x = code within the code segment.

- X = (condition-based lines) + simple lines + looped lines

- X = (1 * 8/10 + 1 * 3 * 8/10) + ( 2 * 8/10 ) + 0

- X = 3.2

- CL = ½ * 3.2 = 1.6

- Code Impact = 1.6 + 2 * 8/10 = 3.2

Consider for the loop shown in Figure-6). Here, the impact of this code depends on the following:-
a) The number of times the loop will be executed i.e. l

b) The impact of the statement, which is a simple statement in this case. If it is a bigger code, say a combination of conditions and loops( nested loops) and more simple statements, then we need to measure it as shown in the next example. We find the impact of each loop, condition and simple statement after which we can add them together to get the impact desired.

In the above example, we will get the impact for the mentioned loop (based on the equations shown for loop-based lines) as:-
    Impact of the for-loop = l * stmts * 0.5
    Where,
     l = variable 'l' which shows how many times the for-loop will execute ( here  l = 10 )
     stmts = Number of statements in the for loop( here it is 1 since there is only one statement)
     Impact of the statement = 0.5
Thus, Impact for the for-loop = 10 * 1 * 0.5 = 5

Consider the entire program code shown in Figure-7. We would have to add the impact of the for-loop ( loop-based lines) and add it to the final code segment impact. Thus we get,
    Code Segment Impact = SL + LL + CL + EL
Where SL = simple lines impact = 1.4, LL = Loop-based lines impact, CL = condition-based lines impact, EL = exception-based lines impact. Here we get the values as:-

Code Segment Impact = 1.4 + 5 + 3.2 = 9.6
- Code Segment Impact = 9.6

Consider the example code at Figure-8) for nested loop condition. We may face conditions where nested loops or nested conditions may come into picture.
To calculate them, we have to follow the same method except for a small change.
Here we replace the variable for representing the statements within the loop or condition by the entire nested loops or condition impact.
For example, in the example above, the entire code has been put into a nested for-loop. Thus, to find the code impact, we shall do the following steps:-

a) The code impact for the following is shown by:-

- Code Impact = SL + CL + LL

- Here, SL and CL are not there outside the loop. Thus, SL = CL = 0

- Next, we know the formula for LL = Counter * Statements * Impact. Here we have

  the same as :-
- Counter = l ( assuming l = 20 in this case) = 20

  Statements = The entire code within the loop has to be measured. Since we have already measured the impact we get it as 11.2
- LL = 20 * 9.6 = 192

- Code Impact = 0 + 0 + 192 = 192

- Thus the code impact for this segment = 192.

Now, if we imagine that the all the examples shown till now were all part of a single program, we would be able to find the overall code segment impact by adding all the impacts that we have found for each of them.
Thus Avg Code Segment Impact = 10 + 5.3 + 3.2 + 9.6 + 192 = 220.1

Next, we find the quality quotient.
Qr = Quality Quotient (on a scale of 1 to 10 given by user). This will include the following aspects [4]:-
1) Security
2) Execution Time
3) User-Friendliness
4) Other software metrics for measuring software quality.
5) Optimum Programming Environment Selection

We will give a measure of 0, 1 or 2 based on the perception of the user of how much is that quality attribute present in the software code. Thus, final answer would be out of 10.

For e.g.) 1/10 till 10/10.

Thus here we will get
1) Security = 1
2) Execution Time = 2
3) User-Friendliness = 0
4) Other software metrics for measuring software quality = 1
5) Optimum Programming Environment Selection = 2

Now adding them we get the quality quotient Qr = 1 + 2 + 0 + 1 + 2 = 6 / 10

Qr = 0.6

Now to get the efficiency, we have:-

Efficiency = (Code Area / Execution Time) * Qr

- Efficiency = (Code Area / Execution Time) * Qr
- Efficiency = ( 220.1 / 88 ) * 6 = 2.50113636363636363636363636364 * 6
- Efficiency = 15.0068181818181818181818181818
- Efficiency = 15.01 ( approximately)

Higher the efficiency, better the code quality is. If we are able to implement our logic with the most minimum number of Lines Of Code(LOC)[3] and/or improve the Quality Quotient attributes, the Efficiency of the code can be improved drastically. This Software Quality Metrics is easy to use, calculate and logically sound. It can be done by any average technical person. This equation again shows the importance of involving the end-user into the software development process. Please note that if there has been a change in the requirements, then we can also find the efficiency lost due to changes by counting the extra or unwanted code written. This can be shown to the client and asked for compensation.

Similarly, the client can ask for the reduction in compensation, if the software that was to be developed is not done on time or within the specified constraints. The client can reduce the impact or the Quality Quotient attributes and reduce the efficiency.

Clearly, this method will give us a very true, clear and accurate idea of the software development process along with more control on the same. This software quality metrics is easy to use and beneficial for anyone, technical or non-technical, who wishes to measure software quality. This metrics can be applied across any language as it depends on the logic rather than the language used for the software.

## II. COMPARISONS/SUPPORT FOR ARGUMENT

1) Halstead Metrics and COCOMO measure the source code and not logic.

2) Function Counts by Albersct's metrics also seems to be excluding the impact of code snippets. Also, this metric measures the input/output i.e. Error messages, user input, etc. It does not consider impact of the code snippets.Morever; cost estimation becomes difficult using this metric as accuracy and logic of the software code are ignored by this metric.

3) Another way of measuring software quality is by the defects it has. This paper measures software quality by its impact and not by defects. Defects Rate measure gives a measure of effectiveness but it isn't efficiency.

## III. ADVANTAGES OF THIS METHOD

1) This method is better than COCOMO as it gives a better estimate of code quality.Every line is measured as per its importance and not by its mere presence as a part of the software code.

2) It measures quality by considering the significance/impact/importance of each code segment. Other well-known quality metrics like COCOMO do not consider such details extensively in the measurements.

3) Halstead Metrics and COCOMO do not measure the code based on its logic, which is the actual essence of the software.

4) The logic of a program/software is implemented differently by different people. One person may implement a simple error-handler in a lotuscript agent or java with 4 lines while another will give a much efficient and extensive error-handler with 5 lines. Both implement the logic of having an error-handler, but with different quality levels and number of lines. Also, certain software requirements are easier to implement in one language while the same would be difficult in another. Thus, there arises the need for the attribute called Impact Factor to measure code quality. The customers get more value for their money as they pay for the logic implemented in the software and not for the number of LOCs in the software. This is because different developers will implement logic in different levels of complexity and number of lines. Better the code quality, higher the impact and thus the cost.

## IV. CONCLUSION

This software quality metric equation measures the logic implemented by the software code and not the lines of code written to implement the logic. This gives a better estimate of the code quality and thus the software cost. The difficult and fuzzy task of cost estimation becomes easier, faster and simpler with this equation. This metric can be easily applied in COCOMO models and other cost estimation models to get the actual software cost. This paper reiterates the importance of quality in the success of any software system by specifying the quality categories for any software. This paper will help the stakeholders of any software to measure and improve the system's performance by using the formula and guidelines described in this paper.

The code segment that is to be selected for getting the impact points based on the logic, solely depends on the individual show understands the system's logic. This categorization of the system into code segments has to be done judiciously and with complete knowledge of the system To get a correct and accurate view of the system, we need to get the code segments and their impacts measured accurately and correctly. It is suggested that the stakeholders from the user-side and the developer sides sit together to perform this activity, thereby helping in obtaining a correct measure of the system's quality. The higher the efficiency, the better will be the software quality. The level of software quality that is expected from the software depends on the developers and the users of the software system. Thus, it is in the hands of these system stakeholders that the quality of the system rests.

APPENDIX

ACL – Access Control List
UI – User Interface
Orgn – organization

ACKNOWLEDGMENT

The author would like to thank his **family and friends** who have motivated and supported him in his work. Also, a special mention of gratitude to **Prof. Uma Srinivasan and Mr. Prashanth Suravjhala,** , who motivated and guided him to persue his research interests. It would not be possible to think and write this much without their motivation and support.

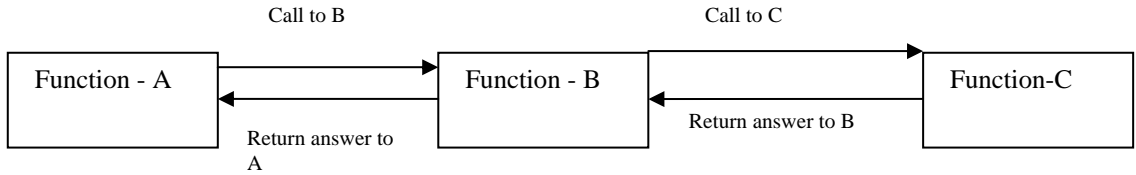

Fig. 3 The Perfect Flow Of Code Execution

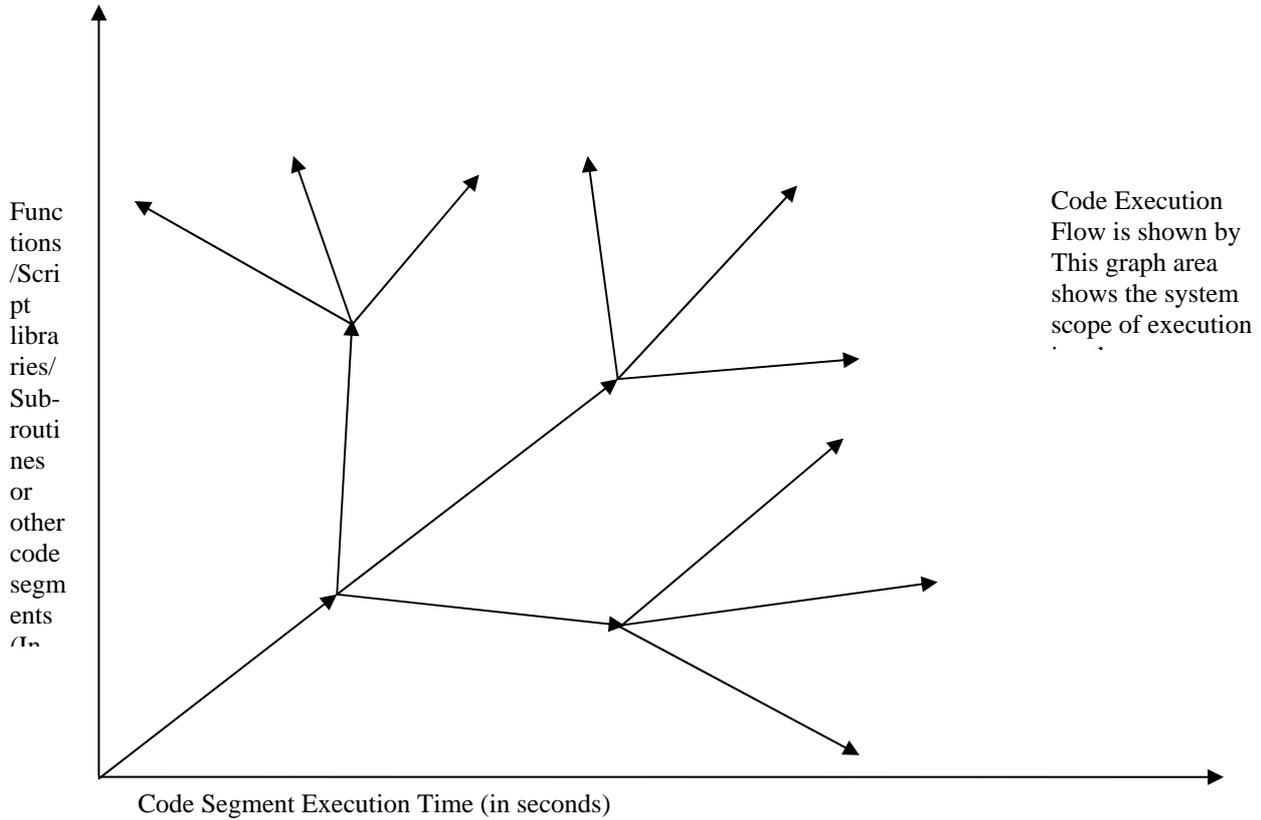

Fig. 4 The Perfect System Flow
(Using Top-Bottom Approach Of Coding)
Note: - For Bottom-Up approach, the arrow heads will be on the opposite sides, moving from outside to inside

Fig. 5 Measuring the simple statements

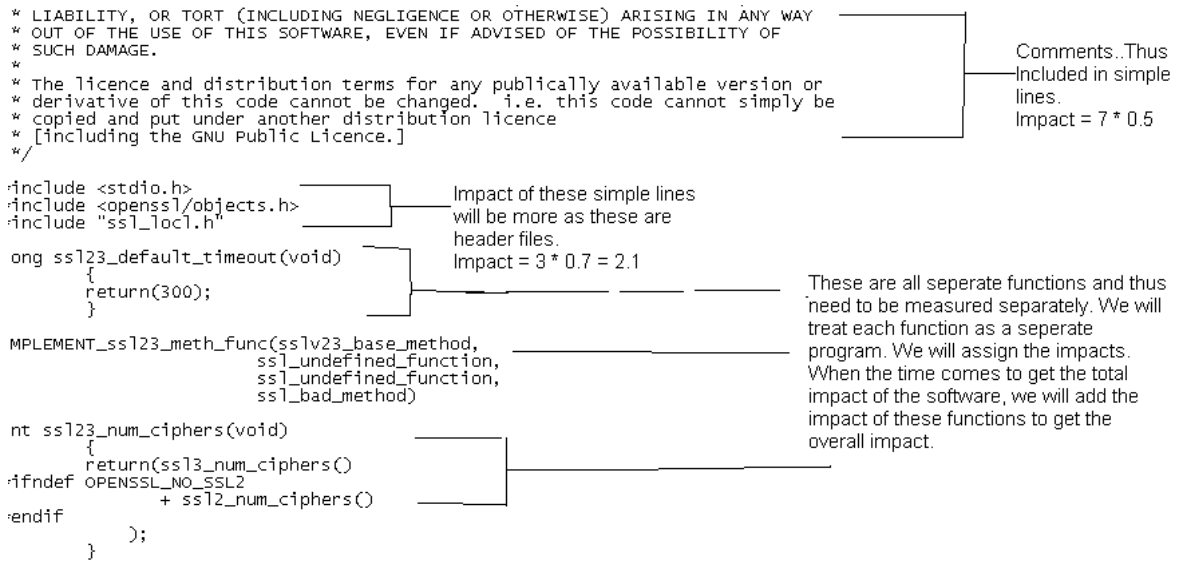

Fig. 6  Measuring the comments

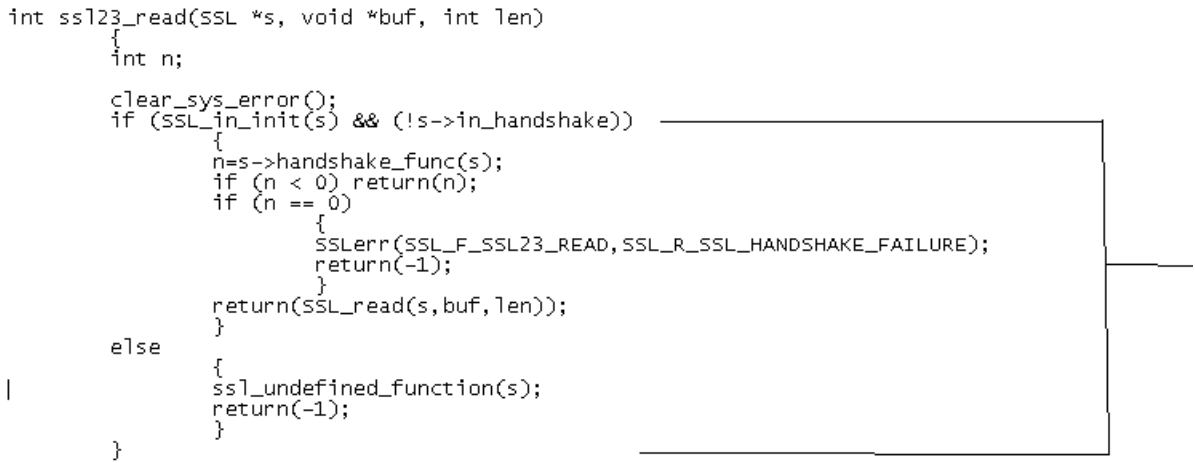

Fig. 7 Measuring the conditional statements

```
for ( i =0 ; i < 1 ; i++ )
    {
        printf("Value of i=%d",i);
    }
```

Fig. 8 Measuring the loop conditions

```
int ssl23_write(SSL *s, const void *buf, int len)
        {
        int n;

        clear_sys_error();
        if (SSL_in_init(s) && (!s->in_handshake))
                {
                n=s->handshake_func(s);
                if (n < 0) return(n);
                if (n == 0)
                        {
                        SSLerr(SSL_F_SSL23_WRITE,SSL_R_SSL_HANDSHAKE_FAILURE);
                        return(-1);
                        }
                return(SSL_write(s,buf,len));
                }
        else
                {
                ssl_undefined_function(s);
                return(-1);
                }

        for ( i =0 ; i < l ; i++ )
                {
                        printf("Value of i=%d",i);
                }
        }
```

Fig. 9 The entire code of the program.

```
int ssl23_write(SSL *s, const void *buf, int len)
{
        for ( i =0 ; i < l ; i++ )
        {
                int n;
                int l = 10 ;
                clear_sys_error();
            if (SSL_in_init(s) && (!s->in_handshake))
             {
                n=s->handshake_func(s);
                if (n < 0) return(n);
                if (n == 0)
                        {
                        SSLerr(SSL_F_SSL23_WRITE,SSL_R_SSL_HANDSHAKE_FAILURE);
                        return(-1);
                        }
                return(SSL_write(s,buf,len));
        }
        else
        {
                ssl_undefined_function(s);
                return(-1);
        }
        for ( i =0 ; i < l ; i++ )
                {
                        printf("Value of i=%d",i);
                }
        }
}
```

Fig. 10 Nested loop condition